# Investigation of the energy levels and the structure of the different states of the $^{24}Mg$ nucleus

Sahar Aslanzadeh[1*], Mohammad Reza Shojaei[1] and Ali Asghar Mowlavi[2]

[1] Departement of Physics, Shahrood University of Technology, Shahrood, Iran
[2] Physics department, Hakim Sabzevari University, Sabzevar, Iran

Correspondence should be addressed to Firstname B. Lastname; s.aslanzadeh@yahoo.co.uk

**Abstract**

In this work, the $^{24}Mg$ nucleus is considered in the cluster model by solving the Schrodinger and Klein- Gordon equations from the Nikiforov- Uvarov (NU) method. A local potential is used for these two equations that is compatible with the Hafstad-Teller potential. By substituting this potential in the Schrodinger and Klein- Gordon equations, the energy levels and wave functions are obtained from NU method. We obtain a first order equation in terms of E for the Schrodinger equation and a fourth-order equation in terms of $E_R$ for the Klein-Gordon equation. Therefore, the obtained equation from the Klein-Gordon equation has both the real solutions and imaginary solutions that we consider only the real solutions for this equation. For more accuracy, the spin-orbit and tensor potentials are added to the central potential as the perturbed terms and the first-order correction value of the energy levels is calculated. These energy levels are used for $^{24}Mg$ nucleus and the good results are obtained in agreement with the experimental data. Of course, one can see that the calculated results from the Schrodinger and Klein-Gordon equations, i.e. non-relativistic and relativistic respectively, are nearly the same. By using the Coulomb repulsive potential, it is shown that the structure of the ground state and the first excited state of this nucleus is as octahedral and the configuration of the alpha particles for the second excited state is as pentahedral, i.e. as $^{12}C + ^{12}C$. Finally, we obtain the critical point for the calculation of maximum energy of this nucleus in the cluster model.

## 1- Introduction

One of the successful models for the consideration of the light nuclei is the cluster model that the best cluster for these nuclei is the alpha particle because this particle has the high binding energy. This model is concluded from this fact that in some nuclei, the alpha particle is radiated. In recent years, the important works have been done about the nuclei with $A = 4n$ that n is the number of the alpha particles like nuclei $^8Be$, $^{16}O$, $^{24}Mg$,... [1, 2, 3, 4, 5]. These nuclei in the cluster model have the different structures [6, 7, 8]. For example, the ground state structure of $^{16}O$ nucleus is as tetrahedral [9, 10, 11] and probably the excited states of this nucleus have the configuration of square, linear chain or non-localized gas [10, 11, 12]. The $^{12}C$ nucleus, composed of three alpha particles, in the ground state is as the linear chain or triangle *alpha* configuration [13, 14, 15, 16]. These nuclei in the ground state have not the cluster structure and with the absorption of some energy will be excited to the cluster configuration [4]. For example, the $^{12}C$ nucleus after absorbing the energy of 2.27 MeV, is excited to three alpha clusters. Now, the $^{24}Mg$ nucleus appears as the two clusters of $^{20}Ne + \alpha$ after taking the energy of 9.32 MeV and by increasing the energy and the absorption of the energy of 13.93 MeV goes into the two cluster configuration of $^{12}C + ^{12}C$ [17]. This nucleus will have the structure composed of the six alpha particles with the excited energy of 24.48 MeV[17].

---

[*] Correspondence: s.aslanzadeh@yahoo.co.uk

In this paper, the $^{24}Mg$ nucleus is considered as a system composed of the six alpha particles. Also, the structure of this nucleus for the different states is studied according to the strength of Coulomb repulsion. In the cluster model, this nucleus has 12 bonds that determine the energy levels of the different states. So, the article goes as follows: in Sec. II-A by using a local potential, the Schrodinger equation is solved from the NU method and the energy levels and the wave functions are gotten. In Sec. II-B the Klein-Gordon equation is solved with this potential from the NU method and again the energy levels are obtained. In Sec. III first the coefficients of the local potential are determined by fitting this potential with the experimental data. In this section, we calculate the energy levels of $^{24}Mg$ nucleus from the previous section and compare them with the experimental data. Also, by using the Coulomb potential, we determine the structure of the several states of this nucleus according to the cluster model. Finally, in Sec. IV the final conclusions are written for this paper.

## 2- the calculation of the energy levels from the NU method

### 2.1 the Schrodinger equation

In this section, the Schrodinger equation is solved and the energy levels are obtained from the NU method. We use a local potential, compatible with the Hafstad-teller potential, that is as[18]:

$$V_C(r) = V_N(r) + V_C(r) = -\frac{V_A}{r} e^{-\frac{r^2}{\alpha_A^2}} + \frac{V_R}{r^2} e^{-\frac{r^2}{\alpha_R^2}} + \frac{V_k}{r}, \qquad (1)$$

where $V_A = V_{0A}\alpha_A$, $V_R = V_{0R}\alpha_R^2$ and $V_k$ are the attraction section, the repulsive section and the strength of the Coulomb's repulsion, respectively. The strength of the Coulomb repulsion is calculated from the Coulomb potential. In above potential, $V_N(r)$ is the local nuclear potential between the alpha particles. In this potential, the repulsive term is because of the Pauli exclusion principle and the attraction term at large distances is for the Van Der Waals potential[18]. In this work, the internal structure of the alpha particles is ignored[19, 20, 21], but the effects of the Pauli exclusion principle have been considered by the repulsive term. So the radial section of the Schrodinger equation is written as:

$$R''(r) + \frac{2}{r}R'(r) + \frac{2\mu}{\hbar^2}\left[E - V(r) - \frac{\hbar^2 l(l+1)}{2\mu r^2}\right]R(r) = 0, \tag{2}$$

where $\mu$, l, E and $R(r)$ are the reduced mass of clusters, the quantum number of angular momentum, the total energy of the system and the radial section of the wave function, respectively. The above equation by substitution of $v(r)$ from Eq. (1) can be written as:

$$\frac{d^2 R(r)}{dr^2} + \frac{2}{r}\frac{dR(r)}{dr} + \frac{1}{r^2}\left[P_1 r e^{-\frac{r^2}{\alpha_A^2}} - P_2 e^{-\frac{r^2}{\alpha_R^2}} - P_3 r - P_0 + K^2 r^2\right]R(r) = 0, \tag{3}$$

where the parameters are as:
$$P_1 = \frac{2\mu V_A}{\hbar^2}, P_2 = \frac{2\mu V_R}{\hbar^2}, P_3 = \frac{2\mu V_C}{\hbar^2}, P_0 = l(l+1), k^2 = \frac{2\mu E}{\hbar^2}. \tag{4}$$

Because we can not solve Eq.(3) from the analytical methods, therefore we must consider approximation for the exponential functions in this equation. So, for performing this work, we should understand the interaction between the alpha particles. Since the binding energy is proportioal to the mass number A , not A(A-1), so the saturation property in the nucleon-nucleon interaction is considered in the cluster-cluster potential too. At the close distances r < 1fm due to the exchange forces between clusters, there is the short-range repulsive forces. At middle distances, the nuclear attractive force between clusters has been considered in our potential. the potential between clusters at the farther distance is under the long-range coulomb potential. This topic is seen in all cluster-cluster interactions Therefore, for using from NU method, the exponential functions in Eq.(3) are approximated up to the degree at most two of r. Now, by simplifying Eq.(3) and approximating the exponential functions up to degree at most two of r, we will have:

$$\frac{d^2 R(r)}{dr^2} + \frac{2}{r}\frac{dR(r)}{dr} + \frac{1}{r^2}\left[\left(\frac{P_2}{\alpha_R^2} + k^2\right)r^2 + (P_1 - P_3)r - (P_2 + P_0)\right]R(r) = 0. \tag{5}$$

Therefore, with defining the parameters of $\alpha, \beta$ and $\gamma$ as:

$$\alpha = -\left(\frac{P_2}{\alpha_R^2} + k^2\right) \cdot \beta = P_1 - P_3 \cdot \gamma = P_2 + P_0, \tag{6}$$

and the change of the variable $r \to s$ and the equivalence $\psi(r) \equiv R(r)$, Eq. (5) is written as:

$$\frac{d^2 \psi(s)}{ds^2} + \frac{2}{s}\frac{d\psi(s)}{ds} + \frac{1}{s^2}[-\alpha s^2 + \beta s - \gamma]\psi(s) = 0. \tag{7}$$

By using Nikiforov-Uvarov (NU) method [22], the above equation can be solved and the energy eigenvalues are gotten as:

$$E = -V_{0R} - \frac{\frac{2\mu}{\hbar^2}(V_{0A}\alpha_A - V_C)^2}{\left(1+2n+\sqrt{(1)+4\left[\frac{2\mu V_{0R}\alpha_R^2}{\hbar^2}+l(l+1)\right]}\right)^2}, \tag{8}$$

that is a first-order equation in terms of E that has the real solutions. The above equation is the energy eigenvalues of nuclei composed of two alpha clusters. Also, the wave function of this equation is [22]:

$$R(r) = \sqrt{\frac{n!(2\sqrt{\alpha})^{\sqrt{1+4\gamma}+2}(2n+\sqrt{1+4\gamma}+1)}{(n+\sqrt{1+4\gamma})!}} e^{-\sqrt{\alpha}r} r^{\frac{1}{2}(\sqrt{1+4\gamma}-1)} L_n^{\sqrt{1+4\gamma}}(2\sqrt{\alpha}\,r) \ . \tag{9}$$

where $L_n^{(\alpha)}$ is the generalized Laguerre polynomials. Now, for more precision, the spin-orbit and tensor terms are added to the central potential as a perturbed potential and the first-order energy shift is shown by $E_n^1$. First, the perturbed potential is written as:

$$V_p(r) = V_{L.S}(r)L.S + V_T(r)S_{12} \ . \tag{10}$$

Where the first term is the spin-orbit potential and the second term is the tensor potential. The first-order correction value of the energy levels that is equal to the expected value of $V_p$ in the $|n>$ state, is as:

$$E_n^{(1)} = < n|V_{L.S}(r)L.S + V_T(r)\hat{S}_{12}|n> =$$
$$\int \psi_n^{(0)*}(r)\big(V_{L.S}(r)L.S + V_T(r)\hat{S}_{12}\big)\psi_n^{(0)}(r)r^2\,dr, \tag{11}$$

where $\psi_n^{(0)}(r)$ is the unperturbed wave function. Therefore, by $\psi_n^{(0)}(r) \equiv R(r)$ and Eq. (30), value of $E_n^{(1)}$ can be calculated[23]..The materials and methods section should contain sufficient detail so that all procedures can be repeated. It may be divided into headed subsections if several methods are described.

## 2.2 the Klein- Gordon equation

The Klein-Gordon equation for a spin-0 particle with the scalar potential of $S(r;\theta;\phi)$ and the vector potential of $V(r;\theta;\phi)$ is written as [24, 25]:

$$\{P^2 - (V(r,\theta,\phi) - E_R)^2 + (V(r,\theta,\phi) + \mu c^2)^2\}\psi(r,\theta,\phi) = 0 \tag{12}$$

where μ, $E_R$, S and V are the rest mass, the relativistic energy, the scalar and vectorial potential, respectively. The above equation can be simplified

as:

$$\nabla^2 + \frac{1}{\hbar^2 c^2}\{(V(r,\theta,\phi) - E_R)^2 - (V(r,\theta,\phi) + \mu c^2)^2\}\psi(r,\theta,\phi) = 0. \tag{13}$$

With selecting $S(\mathbf{r}) = \mathbf{V}(\mathbf{r})$ for the above equation, the low equation is gotten:

$$\nabla^2\psi(r) + \frac{1}{\hbar^2 c^2}\{-2V(r)(E_R + \mu c^2) + E_R^2 - \mu^2 c^4\}\psi(r) = 0 \tag{14}$$

The radiation section of the above equation for the D-dimensional position vector **r** is as:

$$\frac{1}{r^{D-1}}\frac{d}{dr}(r^{D-1}\frac{dR(r)}{dr}) + \frac{1}{\hbar^2 c^2}\{-2V(r)(E_R + \mu c^2) + (E_R^2 - \mu^2 c^4) - \frac{\hbar^2 c^2 l(l+D-1)}{r^2}\}R(r) = 0 \quad (15)$$

With simplifying the above equation, one can obtain

$$\frac{d^2 R(r)}{dr^2} + \frac{D-1}{r}\frac{dR(r)}{dr} + \frac{1}{\hbar^2 c^2}\{\alpha_2(\alpha_1 - 2V) - \frac{\hbar^2 c^2 \lambda_l}{r^2}\}R(r) = 0 \quad (16)$$

where $\alpha_1 = E_R - \mu c^2$, $\alpha_2 = E_R + \mu c^2$ and $\lambda_l = l(l + D - 2)$. By using the potential of Eq.(1) into the above equation and simplifying this equation and with approximating the exponential functions up to degree at most two of r, we get

$$\frac{d^2 R(r)}{dr^2} + \frac{D-1}{r}\frac{dR(r)}{dr} + \frac{1}{r^2}\{\left(\frac{2\alpha_2 V_R}{\alpha_R^2 \hbar^2 c^2} + \frac{\alpha_1 \alpha_2}{\hbar^2 c^2}\right)r^2 + \frac{2\alpha_2(V_{AC}-V_C)}{\hbar^2 c^2}r - (\frac{2\alpha_2 V_R}{\hbar^2 c^2} + \lambda_l)\}R(r) = 0. \quad (17)$$

The above equation with the change of the variables:

$$\acute{\alpha} = -\left(\frac{2\alpha_2 V_R}{\alpha_R^2 \hbar^2 c^2} + \frac{\alpha_1 \alpha_2}{\hbar^2 c^2}\right), \acute{\beta} = \frac{2\alpha_2(V_{AC}-V_C)}{\hbar^2 c^2}, \acute{\gamma} = \left(\frac{2\alpha_2 V_R}{\hbar^2 c^2} + \lambda_l\right), \psi(r) \equiv R(r) \quad (18)$$

for three dimensions, is written as

$$\frac{d^2 R(r)}{dr^2} + \frac{2}{r}\frac{dR(r)}{dr} + \frac{1}{r^2}\{-\acute{\alpha}r^2 + \acute{\beta}r - \acute{\gamma}\}R(r) = 0. \quad (19)$$

Now by using NU method [22], the low equation is gotten:

$$\acute{\alpha} = \frac{\acute{\beta}^2}{\left(1+2n+\sqrt{1+4\acute{\gamma}}\right)^2}. \quad (20)$$

Therefore, by substituting the parameters of Eq. (16) in the above equation, somebody can get:

$$\frac{2\alpha_2 V_R}{\alpha_R^2 \hbar^2 c^2} + \frac{\alpha_1 \alpha_2}{\hbar^2 c^2} = \frac{\frac{4\alpha_2^2(V_{AC}-V_C)^2}{\hbar^4 c^4}}{\left(1+2n+\sqrt{(1)+4\left(\frac{2\alpha_2 V_R}{\hbar^2 c^2}+\lambda_l\right)}\right)^2}. \quad (21)$$

With respect to the parameters of $\alpha_1$ and $\alpha_2$, the above equation is converted to the following quartic equation for calculating $E_R$:

$$64V_R^2 E_R^4 + \{(64\lambda_l + 16(1) - 16(1+2n)^2)\hbar^2 c^2 V_R + \frac{256 V_R^3}{\alpha_R^2} - 64V_R(V_A - V_C)^2\}E_R^3 +$$

$$\{[(1+2n)^4 + (1)^2 + 16\lambda_l^2 + 8(1)\lambda_l - 2(1+2n)^2((1) + 4\lambda_l)]\hbar^4 c^4 + (16\lambda_l + 16(1) -$$

$$16(1+2n)^2)\left(4\frac{V_R^2 \hbar^2 c^2}{\alpha_R^2} - \mu\hbar^2 c^4 V_R\right) + \frac{256 V_R^4}{\alpha_R^4} + \frac{256 V_R^3}{\alpha_R^2}\mu c^2 - 128\mu^2 c^4 V_R^2 + 16(V_A - V_C)^2 -$$

$$8\hbar^2 c^2 (V_A - V_R)^2 ((1) + (1+2n)^2 + 4\lambda_l) - \frac{128 V_R^2 (V_A - V_R)^2}{\alpha_R^2} - 64V_R \mu c^2 (V_A - V_R)^2\}E_R^2 +$$

$$\left\{\left[\left((1+2n)^4+(1)^2+16\lambda_l^2+8(1)\lambda_l-2(1+2n)^2\left((1)+4\lambda_l\right)\right)\right]\left(4\frac{V_R\hbar^4c^4}{\alpha_R^2}-\mu\hbar^4c^6\right)+\right.$$
$$(16\lambda_l+16(1)-16(1+2n)^2)\left(4\frac{\hbar^4c^4V_R^3}{\alpha_R^4}-\mu^2\hbar^4c^6V_R\right)+\frac{512V_R^2\mu c^2}{\alpha_R^4}-\frac{256V_R^3\mu^2c^4}{\alpha_R^2}+32(V_A-V_R)^2\mu c^2-\frac{16\hbar^2c^2V_R(V_A-V_R)^2}{\alpha_R^2}\left((1)+(1+2n)^2+4\lambda_l\right)-\frac{256\mu c^2V_R^2(V_A-V_R)^2}{\alpha_R^2}+64V_R\mu^2c^4(V_A-V_R)^2\right\}E_R+\left\{\left[(1+2n)^4+(1)^2+16\lambda_l^2+8(1)\lambda_l-2(1+2n)^2\left((1)+4\lambda_l\right)\right]\left(4\frac{V_R^2\hbar^4c^4}{\alpha_R^4}-\right.\right.$$
$$4\frac{V_R\mu\hbar^4c^6}{\alpha_R^2}+\mu^2\hbar^4c^8\right)+4(64\lambda_l+16(1)-16(1+2n)^2)\left(\frac{V_R^3\hbar^4c^4}{\alpha_R^4}-\frac{V_R^2\hbar^2\mu^2c^6}{\alpha_R^2}-\hbar^2\mu^3c^8V_R\right)+$$
$$\frac{256V_R^2\mu^2c^4}{\alpha_R^4}+16(V_A-V_R)^2\mu^2c^4+64V_R^2\mu^4c^8+$$
$$\left(8(V_A-V_R)^2\hbar^2c^8\mu^2-\frac{16\mu\hbar^2c^4V_R(V_A-V_R)^2}{\alpha_R^2}\right)\left((1)+(1+2n)^2+4\lambda_l\right)-\frac{128\mu^2c^4V_R^2(V_A-V_R)^2}{\alpha_R^2}+$$
$$64V_R\mu^3c^6(V_A-V_R)^2\bigg\}=0 \qquad (22)$$

Therefore, the above equation is as $aE_R^4+bE_R^3+cE_R^2+dE_R+e=0$ that by determining the parameters of a, b, c, d, and e, the value of $E_R$ can be calculated for this equation. The above equation is an equation of the fourth degree in terms of $E_R$. This equation has both the real solutions and the imaginary solutions that we only consider the real solutions for this equation.

### 3- Example, Discussion, and Results

In this section, we consider the nucleus of $^{24}Mg$ that has six alpha particles. The energy levels in Eqs. (8) and (21) have been obtained for two body systems like $^8Be$. So, these equations will change because the $^{24}Mg$ nucleus consist of six alpha particles. This nucleus can be constructed by a variety of the geometric arrangements of six alpha particles. Therefore, it should determine which one of the $\alpha$ cluster configurations in the $^{24}Mg$ nucleus has the lowest energy, i.e. the ground state of this nucleus. In recent years, the several works have been done about the structure of the different states of the $^{24}Mg$ nucleus [26, 27]. In this section, the ground state structure of this nucleus is determined by the Coulomb's interaction between the alpha particles. Therefore, for simplifying the consideration of this six-particle nucleus, the Jacobi coordinates are used with coordinates of $\xi_1;\xi_2;\xi_3;\xi_4;\xi_5$, and R as:

$$\xi_1=\frac{r_1-r_2}{\sqrt{2}},\xi_2=\frac{r_1+r_2-2r_3}{\sqrt{6}},\xi_3=\frac{r_1+r_2+r_3-3r_4}{\sqrt{12}},\xi_4=\frac{r_1+r_2+r_3+r_4-4r_5}{\sqrt{20}},$$
$$\xi_5=\frac{r_1+r_2+r_3+r_4+r_5-5r_6}{\sqrt{30}},R=\frac{r_1+r_2+r_3+r_4+r_5+r_6}{\sqrt{6}} \qquad (23)$$

Where $r_1,r_2,r_3,r_4,r_5$ and $r_6$ are the relative positions of the six alpha particles in the $^{24}Mg$ nucleus. Now, since the nuclei are almost spherical, it is better to use super-spherical variables. For a N-particle system, we have 3N super-spherical coordinates that 2N coordinates of them are the polar angles, N-1 coordinates are the super-angels, and the last remaining coordinate is super-radius x. Therefore, we introduce the hyper-angles $\theta_1,\theta_2,\theta_3,\theta_4,\theta_5$, and the hyper-radius quantity x as [28, 29, 30]:

$$x=\left(\sum_{i=1}^N\xi_i^2\right)^{\frac{1}{2}}, \qquad \theta_i=\tan^{-1}\left(\frac{\left(\sum_{i=1}^N\xi_i^2\right)^{\frac{1}{2}}}{\xi_{i+1}}\right) \qquad (24)$$

where $i = 1, 2, 3, ... N - 1$ and N is the particle number. The Schrodinger and Klein-Gordon equations respectively in Eqs. (7) and (19) for the N-Body systems in terms of the hyper-radius quantity x are as:

$$\frac{d^2\psi(x)}{dx^2} + \frac{2}{x}\frac{d\psi(x)}{dx} + \frac{1}{x^2}[-\alpha x^2 + \beta x - \gamma]\psi(x) = 0 \quad (25)$$

and

$$\frac{d^2\psi(x)}{dx^2} + \frac{2}{x}\frac{d\psi(x)}{dx} + \frac{1}{x^2}\{-\acute{\alpha} x^2 + \acute{\beta} x - \acute{\gamma}\}\psi(x) = 0 \quad (26)$$

where D=3N-3 and N is the number of particles. Now, by solving Eqs. (25) and (26) from the NU method [22], we obtain the energy levels as:

$$E = -V_{0R} - \frac{\frac{2\mu}{\hbar^2}(V_{0A}\alpha_A - V_C)^2}{\left(1+2n+\sqrt{(2-D)^2+4\left[\frac{2\mu V_{0R}\alpha_R^2}{\hbar^2}+l(l+1)\right]}\right)^2} , \quad (27)$$

and

$$\frac{2\alpha_2 V_R}{\alpha_R^2 \hbar^2 c^2} + \frac{\alpha_1 \alpha_2}{\hbar^2 c^2} = \frac{\frac{4\alpha_2^2(V_{AC}-V_C)^2}{\hbar^4 c^4}}{\left(1+2n+\sqrt{(2-D)^2+4\left(\frac{2\alpha_2 V_R}{\hbar^2 c^2}+\lambda_l\right)}\right)^2} \quad (28)$$

for the Schrodinger and Klein- Gordon equation, respectively. Therefore, for the N-particle systems, the wave function of Eq. (9) and Eqs. (8), (21) and (22) change by substituting the term of $(D-2)^2$ instead of number 1 inside parentheses. For calculating the energy levels of this nucleus, the parameters of Eq. (1) must be determined. So, we first obtain the strength of the Coulomb repulsion, i.e. $V_k$, for this nucleus. The $^{24}Mg$ nucleus has 12 bonds in the alpha cluster model. Therefore, this nucleus has the different configurations like the linear chain state [18], the octahedral structure [19], and pentahedral or similar to $^{12}C + ^{12}C$ state [20]. First, the ground state structure of this nucleus is considered as octahedral that has 12 bonds [20]. Therefore, for this structure, the effective value of $V_k$ in the Coulomb potential $V_C = \frac{V_k}{r}$ is equal 69.07MeV.fm. Also, the reduced mass $\mu$ for a six-body system is about $\frac{5}{6}$ of the mass of the one of particles. So, $\mu$ is equal 3121.2 $\frac{MeV}{c^2}$, respectively. Also, by fitting the ground state and the first excited state of the $^{24}Mg$ nucleus with the experimental data, the parameters of $V_{0A}, V_{0R}, \alpha_A$ and $\alpha_R$ are equal 100 MeV, 194.47 MeV, 2 fm, and 2.13 fm, respectively [31, 32, 33, 34, 35]. Finally, by putting the values of these parameters in Eqs. (27) and (28), the energy levels of the $^{24}Mg$ nucleus from the two equations of Schrodinger and Klein-Gordon are obtained (see Table below) along with the Data experimental that are available in Ref. [36].

Table: the energy levels of nucleus $^{24}Mg$

| Levels | $E_{cal}^S(MeV)$ | $E_{cal}^R(MeV)$ | $E_{cal}^S + E_n^{(1)}$(MeV) | $E_{exp}(MeV)$[36] | $J^P, T, L, S$ |
|---|---|---|---|---|---|
| Ground state | -197.62 | -197.76 | -197.62 | -198.26 | $0^+, 0,0,0$ |
| First excited state | -197.16 | -197.33 | -196.84 | -197.1 | $2^+, 0, 2, 1$ |
| Second excited state | -197.00 | -197.12 | -196.23 | -194.11 | $4^+, 0,4,1$ |

From this Table, one can see that the gotten results of these two equations are roughly the same. The results of this Table have been calculated by assuming that the structure of $^{24}Mg$ in any state is as Octahedral. From this Table, we understand that the ground state has the good agreement with the experimental data and the calculated relativistic value for this state is closer to the experimental data. In this state, because the spin is equal to zero, so $E_n^1 = 0$. Therefore, we conclude that the ground state of this nucleus has the octahedral configuration.

Now, we consider the first excited state by comparing between the obtained results and the experimental data. In this state, the value $E_{cal}^S + E_n^1$ is equal to -196.84 MeV that is very close to the experimental data. Therefore, the structure of this state of the $^{24}Mg$ nucleus in the cluster model is as octahedral, i.e. like the ground state. Of course, because the energies of these two levels experimentally are near together, so it is expected that these two states will have the same structure. Now, if one pays attention to the second excited state from Table, it is seen that there is a notable difference between the calculated results and the experimental data about 2.12 MeV. So anyone can understand that the octahedral structure for this state is not suitable and should consider the other configurations for this state. First, if we suppose that the structure of this state is as pentahedral, i.e. $^{12}C + ^{12}C$ with 12 bands, then $V_C = 6(6 + \sqrt{2})e^2$ that is equal to 64.05 MeV.fm. The value of $E_{cal}^S$ with this $V_C$ is -197.11 MeV that it's different with the octahedral structure is very small about 0.1 MeV. But in this structure, the value of $E_{cal}^S + E_n^1$ is -195.08 MeV that is nearer to the experimental data.

Now we consider the structure of the linear chain that for this configuration $V_C = 30e^2 = 43.197 MeV.fm$. By using this value for $V_k$, the value of $E_{cal}^S$ is equal to -197.98 MeV. By calculating $R(r)$ for this structure from Eq. (9) and substitution R(r) instead of $\psi_n^0(r)$ in Eq. (11), we can get $E_n^1$ for this structure that is equal to 1.86 MeV [22]. Therefore, the value of $E_{cal}^S + E_n^1$ for the linear chain structure is -196.12 MeV. Now, if we pay attention to the obtained results for $E_{cal}^S + E_n^1$, it is gotten that its value for the pentahedral structure is closer to the experimental data. So, somebody can understand that the configuration of the alpha particles for $^{24}Mg$ nucleus in the second excited state is as pentahedral or similar to $^{12}C + ^{12}C$. This is a good result for the $^{24}Mg$ nucleus because we can understand that the higher excited states probably have the decentralized structures like the non-localized gas configuration.

In all above cases, for calculation of the energy levels, the value of $V_k$ is variable. So, for determination of the maximum value of the energy level in Eq. (27), we should make the derivative of E with respect to $V_k$. This derivative is written as:

$$\frac{dE}{dV_k} = \frac{\frac{4\mu}{\hbar^2}(V_{0A}\alpha_A - V_C)}{\left(1 + 2n + \sqrt{(2-D)^2 + 4\left[\frac{2\mu V_{0R}\alpha_R^2}{\hbar^2} + l(l+1)\right]}\right)^2} \qquad (29)$$

If we set the above equation equal to zero, then the critical points are obtained that the value of E is maximum at them. For the above equation, the critical point is as:

$$V_k = V_{0A}\alpha_A \qquad (30)$$

where according to the value of above parameters, $V_k$ is about 200 MeV.fm. Therefore, at this $V_k$, the value of E is maximum. One can simply understand that in this state, the alpha particles are very close together. Also, we can conclude that this state is unstable because the distance between alpha particles is very small comparing with the ground state.

## 4-Conclusions

In this paper, the $^{24}Mg$ nucleus has been considered in the cluster model. The Schrodinger and Klein-Gordon equations were solved for the many-alpha systems with a local potential from NU method and the energy levels and the wave functions were obtained. We got a first-order equation in terms of E for the Schrodinger equation and a fourth-order equation in terms of $E_R$ for the Klein-Gordon equation. So, we concluded that the obtained equation from the Klein-Gordon equation has both the real solutions and imaginary solutions. We considered only the real solutions for this equation. We exerted the obtained results for the nucleus of $^{24}Mg$ and concluded that according to the above Table, the results of these two equations are nearly the same. Now, by comparing the calculated energy with the experimental data, it was gotten that the structure of the ground state and the first excited state of the $^{24}Mg$ nucleus in the cluster model is as the octahedral. Finally, it was obtained that the second excited state of the $^{24}Mg$ nucleus in the cluster model has the structure of pentahedral or similar to $^{12}C + ^{12}C$ because the calculated results for this configuration were closer to the experimental data. Finally, we obtained the state that its energy is Maximum. Also, we concluded that according to the value of $V_k$ for this state, the alpha particles are very close together and therefore this state is unstable.

١